\documentclass[11pt]{article}
\usepackage[letterpaper,margin=1in]{geometry}
\usepackage[T1]{fontenc}
\usepackage[utf8]{inputenc}
\usepackage{lmodern,microtype}
\usepackage{amsmath,amssymb,amsthm,mathtools,bm}
\usepackage{booktabs,tabularx,array,multirow}
\usepackage{graphicx}
\usepackage{xcolor}
\usepackage{enumitem}
\usepackage{natbib}
\usepackage{hyperref}
\usepackage[nameinlink,capitalise,noabbrev]{cleveref}
\usepackage{fancyhdr,lastpage}
\usepackage{titlesec}

\definecolor{siamblue}{HTML}{17365D}
\definecolor{teal}{HTML}{007C83}
\hypersetup{
  hidelinks,
  pdftitle={A Second Order Transport Scheme for Dynamic Portfolio Subspaces under a Varying Risk Metric},
  pdfauthor={Alejandro Rodriguez Dominguez},
  pdfsubject={Research preprint on geometric transport in portfolio subspaces},
  pdfkeywords={portfolio representation, parallel transport, risk metric, Procrustes alignment, geometric integration}
}
\titleformat{\section}{\Large\bfseries}{\thesection.}{0.6em}{}
\titleformat{\subsection}{\large\bfseries}{\thesubsection.}{0.6em}{}
\titleformat{\subsubsection}{\normalsize\bfseries}{\thesubsubsection.}{0.5em}{}
\numberwithin{equation}{section}
\setlist{leftmargin=1.5em}
\allowdisplaybreaks[1]
\newcommand{\proofstep}[1]{\par\addvspace{0.4\baselineskip}\noindent\textit{#1}\enspace}

\newtheorem{theorem}{Theorem}[section]
\newtheorem{proposition}[theorem]{Proposition}
\newtheorem{corollary}[theorem]{Corollary}

\theoremstyle{definition}

\theoremstyle{remark}

\newtheorem{assumption}[theorem]{Assumption}

\newcommand{\R}{\mathbb R}
\newcommand{\Spp}{\mathbb S_{++}}
\newcommand{\cV}{\mathcal V}
\newcommand{\cX}{\mathcal X}
\newcommand{\cE}{\mathcal E}

\newcommand{\ip}[2]{\left\langle #1,#2\right\rangle}
\newcommand{\norm}[2][]{\left\lVert #2\right\rVert_{#1}}

\newcommand{\polar}{\operatorname{polar}}

\newcommand{\im}{\operatorname{im}}

\newcommand{\diag}{\operatorname{diag}}
\newcommand{\Skew}{\operatorname{skew}}
\newcommand{\doi}[1]{\href{https://doi.org/#1}{doi:#1}}
\newcolumntype{Y}{>{\raggedright\arraybackslash}X}

\title{A Second Order Transport Scheme for Dynamic Portfolio Subspaces under a Varying Risk Metric}
\author{Alejandro Rodr\'iguez Dom\'inguez\thanks{Quantitative Analysis Department, Miralta Finance Bank S.A., Madrid, Spain, Department of Computer Science, University of Reading, Reading, United Kingdom. Email: \href{mailto:arodriguez@miraltabank.com}{arodriguez@miraltabank.com}. ORCID: \href{https://orcid.org/0000-0002-2400-1097}{0000-0002-2400-1097}.}}
\date{}

\begin{document}
\maketitle

\begin{abstract}
We analyze the transport of portfolio exposures between moving subspaces under a varying positive definite risk metric. The reference evolution is the induced connection obtained from the ambient connection of minimum norm in fixed asset coordinates. An explicit expansion of risk weighted Procrustes identifies a leading local error that couples metric deformation with covariant motion normal to the subspace. A rank two construction with an exact continuous solution shows that this term can reduce global accuracy to first order even when the metric commutes with its derivative. Geometric mean alignment and a symmetric Procrustes construction cancel the term and yield reversible updates with second order global accuracy. We establish uniform error bounds on nonuniform partitions, characterize the cases in which the Procrustes defect vanishes, and distinguish discretization error from errors in estimated geometry. We also characterize the connection induced by principal square root whitening and its relation to the selected transport. Manufactured examples confirm the predicted defect and convergence rates. A historical geometry stress test then estimates the defect from preceding observations and finds that it predicts the subsequent separation between risk weighted Procrustes and both second order updates, while the two second order schemes remain much closer to each other. Portfolio comparisons separately evaluate target persistence, turnover, and economic implementation.
\end{abstract}




\section{Introduction}

Rolling portfolio construction estimates a sequence of subspaces in which a strategy can take risk, while implementation requires an additional rule for carrying existing exposures between successive estimates. Changes in the subspace, its covariance metric, and the basis returned by an eigensolver all affect the coordinates used to describe a position. A direct comparison of coefficient vectors consequently combines changes in the investment target with changes in its coordinate representation, even before persistence or transaction costs enter the allocation rule.

This issue arises in factor portfolios, reduced rank risk models, constrained response spaces, and rolling principal component systems. Estimation errors and constraints affect portfolio stability \citep{jagannathanma2003,demiguel2009,brodie2009}, while dynamic trading models make the adjustment of existing positions part of the investment decision \citep{garleanu2013}. Dynamic causal portfolio choice provides a related setting in which a moving conditioning representation enters an intertemporal control problem \citep{rodriguez2026dynamic}. Conditional on a supplied sequence of subspaces, the present paper studies the numerical transport of their exposures under a changing risk metric.

Polar alignment, induced connections, and positive definite matrix geometry provide the ingredients for comparing frames and transporting vectors between such subspaces \citep{higham1986,edelman1998,harland2023,pennec2006}. Preservation of endpoint inner products is distinct from consistency with a prescribed continuous connection, as illustrated by propagation in evolving nonorthogonal bases \citep{artacho2017}. The approximation problem therefore requires both a specified connection and an analysis of the discrete update relative to its continuous transport.

For the metric compatible connection selected here, we derive the leading local error of risk weighted Procrustes as a skew matrix coupling metric deformation with motion normal to the current subspace. A commuting metric example with an exact continuous solution establishes the sharpness of the resulting first order global error. Geometric mean alignment cancels this term and gives a reversible second order update, while symmetrization of forward and backward Procrustes provides an alternative of the same order with lower dimensional matrix operations. The analysis also characterizes the constant metric, fixed subspace, and other configurations in which the defect vanishes and risk weighted Procrustes retains second order accuracy.

Bounds for estimated frames and connections distinguish discretization error from uncertainty in the underlying geometry. The convergence analysis assumes smooth paths of fixed rank and uniformly positive metrics, with spectral separation required when an eigenspace perturbation bound is used. Manufactured examples test the error coefficient and its predicted accumulation, while historical portfolio comparisons hold signals, covariance estimates, risk targets, persistence weights, and trading conventions fixed across transport rules. This design separates the turnover and performance effects of target persistence from the smaller differences between the metric based updates. Section 2 positions the analysis within the literature, Section 3 develops the theory, Section 4 presents the numerical results, and Section 5 concludes.

\section{Related Literature}

Portfolio selection and intertemporal investment theory provide the financial setting \citep{markowitz1952,merton1971}. Constraints and regularization can stabilize estimated portfolios \citep{jagannathanma2003,demiguel2009,brodie2009}, while covariance shrinkage and factor based estimation stabilize the risk input \citep{ledoitwolf2004,fan2013}. Dynamic trading with predictable returns and transaction costs studies the choice of an adjustment policy \citep{garleanu2013}. These contributions address target estimation and investment decisions, whereas the present analysis concerns transport between successive supplied subspaces, a numerical component of the representation dynamics considered in \citet{rodriguez2026dynamic}.

Principal angles and perturbation theory describe the separation and estimation of subspaces \citep{bjorckgolub1973,daviskahan1970,stewartsun1990}. The orthogonal Procrustes problem and polar decomposition provide the associated frame alignment \citep{schonemann1966,gower2004,higham1986}, and the geometry of Stiefel and Grassmann manifolds provides connections and computational procedures \citep{edelman1998,absil2008,bendokat2024}. These constructions underpin the endpoint schemes analyzed below.

For parallel transport of induced connections, \citet{littlejohn2022} show that the singular value basis and the parallel transported basis agree through second order but not beyond. \citet{harland2023} develop gauge covariant approximations of induced transport through order four, and \citet{zhang2026} analyze second order discrete polar transport and spectral interpolation of analytic invariant subspaces. Harland's framework covers an arbitrary metric compatible connection after an ambient parallel trivialization along the curve, whose construction requires the ambient propagator. We study the corresponding endpoint problem, in which only the metrics and subspace frames at consecutive observations are available and the ambient differential equation is not solved in advance.

The distinction between preservation of inner products and consistency with a prescribed evolution is developed by \citet[Sec.~III.A.4 and Appendix~K]{artacho2017}, who show that propagation through consecutive L\"owdin orthonormalizations can preserve orthogonality while failing to represent the required evolution. Their basis overlap connection is supplied by the motion of the basis functions. In the present setting, a connection is selected from the asset risk metric, and the discrepancy of the principal square root map is evaluated relative to that choice. The metric derivative term follows the established geometry of evolving metrics \citep{gracia2026}, while the calculation below identifies its effect on the specified portfolio transport.

The matrix geometric mean and the affine invariant geometry of positive definite matrices provide the endpoint isometry used in the geometric mean scheme \citep{moakher2005,pennec2006,bhatia2007}. Alternative geometries lead to different interpolations and transport laws \citep{arsigny2007,thanwerdas2019}, making the choice of connection part of the specification. Matrix function calculus and geometric integration supply the analytical tools \citep{higham2008,hairer2006}. The present comparison builds on these results and on second order polar transport in a fixed metric to determine the additional error introduced when the metric and subspace vary together.

The resulting contribution is an explicit leading error formula for risk weighted Procrustes, an exact example establishing its sharpness, and two endpoint constructions that cancel the error. The formula identifies the interaction responsible for the loss of one order and the conditions under which fixed metric accuracy extends to a varying metric. The square root discrepancy, coordinate transformation laws, and stability estimates characterize the resulting transport and its sensitivity to estimated inputs, while the historical experiments quantify the effects of the alternative updates within a common portfolio implementation.

\section{Mathematical Developments}

\subsection{Risk metric and investable representation}

Let $\cX$ be a smooth finite dimensional state and time manifold and consider the trivial asset weight bundle $\cX\times\R^n$. At $x\in\cX$, let
\begin{equation}\label{eq:metric}
 Q_x\in\Spp^n,
 \qquad
 \ip{y}{z}_{Q_x}=y^\top Q_xz,
\end{equation}
where $Q$ is three times continuously differentiable on the region of interest. Linear constraints are represented by a full column rank matrix $H_x$, so feasible perturbations lie in $\ker H_x^\top$. The statistical or economic model supplies a rank-$r$ representation $\cE_x\subset\R^n$. A local constrained response operator $M_x$ maps represented directions into feasible portfolio directions. The transported object is the investable image
\begin{equation}\label{eq:bundle}
 \cV_x:=\im(M_x|_{\cE_x})\subseteq\ker H_x^\top,
 \qquad \dim\cV_x=p,
\end{equation}
where the rank satisfies $1\leq p\leq n$ and is constant. Thus $\cV\to\cX$ is a smooth subbundle. A local frame $U(x)\in\R^{n\times p}$ is chosen $Q$-orthonormally:
\begin{equation}\label{eq:frame}
 U^\top Q U=I_p,
 \qquad
 P:=UU^\top Q.
\end{equation}
Then $P^2=P$, $P^\top Q=QP$, and $P$ is the $Q$-orthogonal projector onto $\cV_x$. The gauge group is $O(p)$: another frame of the same subspace has the form $U'=UR$ for a smooth $R(x)\in O(p)$.

\subsection{Compatible ambient connections and the selected transport}

In local state coordinates, write an ambient connection as
\begin{equation}\label{eq:ambient}
 D_a y=\partial_a y+A_a y.
\end{equation}
Metric compatibility requires
\begin{equation}\label{eq:compatibility}
 \partial_a Q=A_a^\top Q+QA_a.
\end{equation}

\begin{proposition}[Compatible family and minimum norm selector]\label{thm:ambient}
For every coordinate direction, the compatible coefficients form the affine family
\begin{equation}\label{eq:family}
 A_a=\tfrac12Q^{-1}\partial_aQ+Q^{-1}K_a,
 \qquad K_a^\top=-K_a.
\end{equation}
Among this family, the risk weighted Hilbert--Schmidt functional
\begin{equation}\label{eq:objective}
 \Phi_Q(A_a)=\norm[F]{Q^{1/2}A_aQ^{-1/2}}^2
\end{equation}
has the unique minimizer
\begin{equation}\label{eq:amin}
 A_a^{\min}=\tfrac12Q^{-1}\partial_aQ.
\end{equation}
The selected connection transforms covariantly under constant changes of asset coordinates, although compatibility alone does not imply uniqueness.
\end{proposition}

\begin{proof}
\proofstep{Compatible coefficients.}
Introduce the risk orthonormal coefficients
\[
 B_a=Q^{1/2}A_aQ^{-1/2},
 \qquad
 C_a=Q^{-1/2}(\partial_aQ)Q^{-1/2}.
\]
Multiplying \cref{eq:compatibility} on both sides by $Q^{-1/2}$ gives
\[
 B_a+B_a^\top=C_a
 \quad\Longleftrightarrow\quad
 B_a=\tfrac12C_a+S_a,
 \qquad S_a^\top=-S_a.
\]
Returning to asset coordinates and setting $K_a=Q^{1/2}S_aQ^{1/2}$ therefore yields
\[
 A_a=\tfrac12Q^{-1}\partial_aQ+Q^{-1}K_a,
 \qquad K_a^\top=-K_a,
\]
which is exactly the family in \cref{eq:family}.

\proofstep{Minimum norm and coordinate covariance.}
The symmetric matrix $C_a$ and the skew symmetric matrix $S_a$ are orthogonal in the Frobenius inner product, so
\[
 \Phi_Q(A_a)
 =\bigl\|\tfrac12C_a+S_a\bigr\|_F^2
 =\tfrac14\|C_a\|_F^2+\|S_a\|_F^2.
\]
Consequently, the unique minimizer has $S_a=0$ and is given by \cref{eq:amin}. For a constant nonsingular coordinate change $C$, write
\[
 Q'=C^\top QC,
 \qquad A_a'=C^{-1}A_aC,
 \qquad O=Q^{1/2}C(Q')^{-1/2}.
\]
Since $O^\top O=I_n$, the corresponding coefficients satisfy
\[
 (Q')^{1/2}A_a'(Q')^{-1/2}=O^\top B_aO,
 \qquad \Phi_{Q'}(A_a')=\Phi_Q(A_a).
\]
The minimizer therefore transforms covariantly. This uniqueness comes from the minimization criterion rather than compatibility alone, because when $Q=I_2$ every constant skew symmetric $A_a$, including a nonzero one, satisfies \cref{eq:compatibility}.
\end{proof}

The compatible family and its metric derivative term follow the established geometry of time dependent metrics \citep{gracia2026}. The proposition specifies the selector used by the numerical schemes below through minimization of the instantaneous coefficient norm in economically fixed asset coordinates. This is a connection on the asset bundle, distinct from a Levi-Civita connection on the state manifold, and its stated covariance concerns constant changes of asset coordinates.

\subsection{Projected connection and gauge covariance}

Project the selected ambient connection onto the investable subbundle:
\begin{equation}\label{eq:induced}
 \nabla_a^{\cV}v:=P D_a^{\min}v,
 \qquad v\in\Gamma(\cV).
\end{equation}

\begin{proposition}[Induced connection and gauge law]\label{prop:gauge}
The connection in \cref{eq:induced} is compatible with the restricted risk metric. In a $Q$-orthonormal frame its connection form is
\begin{equation}\label{eq:omega}
 \omega_a
 =U^\top Q(\partial_aU+A_a^{\min}U)
 =U^\top Q\partial_aU+\tfrac12U^\top(\partial_aQ)U,
 \qquad \omega_a^\top=-\omega_a.
\end{equation}
Under $U'=UR$,
\begin{equation}\label{eq:gaugelaw}
 P'=P,
 \qquad
 \omega_a'=R^\top\omega_aR+R^\top\partial_aR.
\end{equation}
Consequently, parallel transported asset space vectors and their risk norms are gauge invariant.
\end{proposition}

\begin{proof}
For sections $v,w\in\Gamma(\cV)$, ambient compatibility and the identities $Pv=v$, $Pw=w$, and $P^\top Q=QP$ give
\begin{align*}
 \partial_a\ip{v}{w}_Q
 &=\ip{D_a^{\min}v}{w}_Q+\ip{v}{D_a^{\min}w}_Q\\
 &=\ip{PD_a^{\min}v}{w}_Q+\ip{v}{PD_a^{\min}w}_Q.
\end{align*}
This proves compatibility of the projected connection. Differentiating the frame constraint gives
\[
 (\partial_aU)^\top QU+U^\top(\partial_aQ)U
 +U^\top Q\partial_aU=0,
\]
so the expression in \cref{eq:omega} satisfies $\omega_a+\omega_a^\top=0$. For the rotated frame $U'=UR$, the projector and the covariant frame derivative are
\begin{align*}
 P'&=URR^\top U^\top Q=P,\\
 D_a^{\min}(UR)&=(D_a^{\min}U)R+U\partial_aR.
\end{align*}
Multiplication by $R^\top U^\top Q$ yields the transformation law
\[
 \omega_a'=R^\top\omega_aR+R^\top\partial_aR.
\]
With $c'=R^\top c$, differentiation of $R^\top R=I_p$ then gives
\[
 dc'+\omega'c'=R^\top(dc+\omega c),
 \qquad U'c'=Uc.
\]
Thus the two coefficient equations describe the same transported asset vector, whose risk norm is consequently independent of the frame.
\end{proof}

\subsection{Continuous transport}

Along a differentiable curve $x(t)$, write $v_t=U(x_t)c_t$ and $\omega_t=\omega_a(x_t)\dot x_t^a$. Parallel transport is
\begin{equation}\label{eq:parallel}
 \dot c_t+\omega_t c_t=0.
\end{equation}
Because $\omega_t$ is skew-symmetric,
\begin{equation}\label{eq:isometry}
 \frac{d}{dt}\norm[Q_t]{v_t}^2
 =\frac{d}{dt}\norm{c_t}^2=0.
\end{equation}

The analysis below concerns differentiable metric and subspace paths observed on a discrete decision grid. Stochastic extensions require a separately specified horizontal lift and are not needed for the consistency or convergence results established here.

\subsection{Portfolio motion and intrinsic rebalancing}

Let a represented portfolio be $\pi_t=\pi_t^0+v_t$, where $\pi^0$ is a feasible anchor and $v_t\in\cV_{x_t}$. Define the normal differential of the frame by
\begin{equation}\label{eq:normal}
 N_a:=(I-P)D_a^{\min}U.
\end{equation}

\begin{proposition}[Orthogonal motion decomposition]\label{thm:motion}
For every differentiable lift $v_t=U_t c_t$,
\begin{equation}\label{eq:decomp}
 D_t^{\min}v_t
 =U_t(\dot c_t+\omega_t c_t)+N_a(x_t)\dot x_t^a c_t,
\end{equation}
and the two terms are $Q_t$-orthogonal. Hence
\begin{equation}\label{eq:pythagoras}
 \norm[Q_t]{D_t^{\min}v_t}^2
 =\norm{\dot c_t+\omega_t c_t}^2
 +\norm[Q_t]{N_a\dot x_t^a c_t}^2.
\end{equation}
At a fixed current vector and state velocity, parallel transport uniquely removes the tangential component and therefore minimizes instantaneous covariant motion. The normal term is invariant under a change of frame.
\end{proposition}

\begin{proof}
The product rule for $v_t=U_tc_t$, followed by the decomposition $I_n=P_t+(I_n-P_t)$, gives
\begin{align*}
 D_t^{\min}v_t
 &=U_t\dot c_t+(D_t^{\min}U_t)c_t\\
 &=U_t\dot c_t+P_t(D_t^{\min}U_t)c_t
   +(I_n-P_t)(D_t^{\min}U_t)c_t\\
 &=U_t(\dot c_t+\omega_tc_t)+N_a(x_t)\dot x_t^ac_t.
\end{align*}
The last term is normal to $\cV_{x_t}$ because
\[
 U_t^\top Q_tN_a
 =U_t^\top Q_t(I_n-P_t)D_a^{\min}U_t=0.
\]
Together with $U_t^\top Q_tU_t=I_p$, this proves the norm decomposition in \cref{eq:pythagoras}. At fixed $c_t$ and state velocity, the only term depending on $\dot c_t$ is $\|\dot c_t+\omega_tc_t\|^2$, whose unique minimum is attained at $\dot c_t=-\omega_tc_t$. Under $U'=UR$, the extra derivative term lies in the represented subspace and is annihilated by $I_n-P$, so
\[
 N_a'=(I_n-P)\bigl((D_a^{\min}U)R+U\partial_aR\bigr)=N_aR.
\]
Since $c'=R^\top c$, it follows that $N_a'\dot x^ac'=N_a\dot x^ac$, proving invariance of the normal asset vector.
\end{proof}

For a differentiable path, define geometric rebalancing energy by
\begin{equation}\label{eq:energy}
 \mathcal E_{\rm geo}[v]
 =\int_0^T\norm[Q_t]{(I-P_t)D_t^{\min}v_t}^2dt.
\end{equation}
The resulting energy is a geometric quantity rather than dollar turnover, which is defined separately on the decision grid. If $\pi_{t_k^-}$ denotes the portfolio immediately before trading, after buy and hold drift since the preceding decision, then
\begin{equation}\label{eq:turnover}
 \operatorname{TO}(\pi)=\frac12\sum_{k=1}^{K}\norm[1]{\pi_{t_k}-\pi_{t_k^-}}.
\end{equation}

\subsection{Endpoint algorithms and the noncommutative obstruction}

Let $(Q_0,U_0)$ and $(Q_1,U_1)$ be consecutive estimates with $U_i^\top Q_iU_i=I_p$. For a nonsingular square matrix $C$, write $\polar(C)=C(C^\top C)^{-1/2}$ for its orthogonal polar factor. The following endpoint rules are compared in the same smooth reference frame, while their implementation permits arbitrary orthogonal choices of basis at each date.

Risk weighted Procrustes first normalizes $U_0$ in the new metric and aligns it with $U_1$:
\begin{equation}\label{eq:rp}
 \widetilde U_0=U_0(U_0^\top Q_1U_0)^{-1/2},
 \qquad
 R_{\rm RP}=\polar(U_1^\top Q_1\widetilde U_0).
\end{equation}

The principal square root map, also used in propagation through L\"owdin orthonormalizations \citep{artacho2017}, is
\begin{equation}\label{eq:root}
 T_{\rm root}=Q_1^{-1/2}Q_0^{1/2}.
\end{equation}
Although it is an exact endpoint isometry, its consistency depends on the connection being approximated.

\begin{theorem}[Principal square root discrepancy]\label{thm:obstruction}
At a given time, write $Q=V\diag(\lambda_1,\ldots,\lambda_n)V^\top$ and $H=V^\top\dot QV$. In the eigenbasis of $Q$, the derivative of \cref{eq:root} is
\begin{equation}\label{eq:rootderivative}
 (\dot T_{\rm root})_{ij}
 =-\frac{H_{ij}}{\sqrt{\lambda_i}(\sqrt{\lambda_i}+\sqrt{\lambda_j})},
\end{equation}
whereas the derivative of the selected minimum norm transport is
\begin{equation}\label{eq:minderivative}
 (-A^{\min})_{ij}=-\frac{H_{ij}}{2\lambda_i}.
\end{equation}
Their difference $K=\dot T_{\rm root}+A^{\min}$ satisfies
\begin{equation}\label{eq:kterm}
 K_{ij}=\frac{H_{ij}(\sqrt{\lambda_j}-\sqrt{\lambda_i})}
 {2\lambda_i(\sqrt{\lambda_i}+\sqrt{\lambda_j})},
 \qquad QK+K^\top Q=0.
\end{equation}
Thus the discrepancy is $Q$-skew-adjoint. The two ambient connections agree if and only if $[Q,\dot Q]=0$, equivalently if $\dot Q$ has no terms between unequal eigenspaces of $Q$.
\end{theorem}

\begin{proof}
For $f(s)=s^{-1/2}$, the divided difference formula for the Fr\'echet derivative \citep[Chap.~3]{higham2008} gives, including the continuous diagonal values,
\[
 f^{[1]}(\lambda_i,\lambda_j)
 =-\frac{1}{\sqrt{\lambda_i\lambda_j}
              (\sqrt{\lambda_i}+\sqrt{\lambda_j})}.
\]
At the initial endpoint $T_{\rm root}=I_n$, so differentiation in the eigenbasis of $Q$ yields
\begin{align*}
 (\dot T_{\rm root})_{ij}
 &=\bigl(Df_Q[\dot Q]Q^{1/2}\bigr)_{ij}\\
 &=f^{[1]}(\lambda_i,\lambda_j)H_{ij}\sqrt{\lambda_j}\\
 &=-\frac{H_{ij}}
 {\sqrt{\lambda_i}(\sqrt{\lambda_i}+\sqrt{\lambda_j})}.
\end{align*}
Subtracting the selected transport coefficient gives
\begin{align*}
 K_{ij}
 &=-\frac{H_{ij}}{\sqrt{\lambda_i}(\sqrt{\lambda_i}+\sqrt{\lambda_j})}
   +\frac{H_{ij}}{2\lambda_i}\\
 &=\frac{H_{ij}(\sqrt{\lambda_j}-\sqrt{\lambda_i})}
 {2\lambda_i(\sqrt{\lambda_i}+\sqrt{\lambda_j})}.
\end{align*}
Because $H$ is symmetric, the metric adjoint relation follows entry by entry from
\[
 (QK+K^\top Q)_{ij}=\lambda_iK_{ij}+\lambda_jK_{ji}=0.
\]
Finally, $K_{ij}$ vanishes precisely when either $H_{ij}=0$ or $\lambda_i=\lambda_j$, and these are also the conditions for
\[
 [\diag(\lambda_1,\ldots,\lambda_n),H]_{ij}
 =(\lambda_i-\lambda_j)H_{ij}=0.
\]
Thus $K=0$ if and only if $[Q,\dot Q]=0$, with no restriction on the choice of basis within a repeated eigenspace.
\end{proof}

This calculation is a specialization of the known distinction between norm preservation and faithful propagation \citep{artacho2017}. Its relation to the usual Euclidean construction can be stated exactly. Set $S=Q^{1/2}$ and $Z=SU$, and define
\begin{equation}\label{eq:whiteningconnection}
 \mathcal B=SA^{\min}S^{-1}-\dot S S^{-1},
 \qquad
 \omega=Z^\top\dot Z+Z^\top\mathcal B Z,
 \qquad \mathcal B^\top=-\mathcal B.
\end{equation}
The principal square root overlap is exactly $Z_1^\top Z_0$, so applying the fixed Euclidean polar method to the whitened frames omits $Z^\top\mathcal B Z$. Moreover, $\mathcal B=SKS^{-1}$ and the resulting coefficient update has the expansion
\begin{equation}\label{eq:projectedrootdefect}
 R_{\rm root}=I_p-h\omega+hU^\top QKU+O(h^2).
\end{equation}
Consequently, its limiting induced connection agrees with the selected one precisely when $U^\top QKU=0$ along the path. Ambient noncommutativity alone does not rule out this projected cancellation, which always holds at rank one. Equations~\eqref{eq:whiteningconnection} and \eqref{eq:projectedrootdefect} follow by differentiating $U=S^{-1}Z$ and taking the skew part of the overlap derivative. They explain why a direct change to whitened coordinates does not remove the choice of connection. Principal square root whitening is tied to the chosen fixed asset coordinates, so its commutation criterion is stated in those coordinates rather than as a congruence invariant condition. The endpoint map associated with the affine invariant geodesic uses established positive definite matrix geometry \citep{moakher2005,pennec2006,bhatia2007}. Define
\begin{equation}\label{eq:geodesicmap}
 G=Q_0^{-1/2}Q_1Q_0^{-1/2},
 \qquad
 T_{\rm geo}=Q_0^{-1/2}G^{-1/2}Q_0^{1/2}.
\end{equation}
Then $T_{\rm geo}^\top Q_1T_{\rm geo}=Q_0$. The coefficient update is
\begin{equation}\label{eq:geoupdate}
 R_{\rm geo}=\polar(U_1^\top Q_1T_{\rm geo}U_0),
 \qquad c_1=R_{\rm geo}c_0.
\end{equation}

\begin{proposition}[Geometric mean form and reversibility]\label{prop:geometricmean}
The standard matrix geometric mean is
\[
 Q_0\#Q_1=Q_0^{1/2}(Q_0^{-1/2}Q_1Q_0^{-1/2})^{1/2}Q_0^{1/2}.
\]
It gives the identities
\begin{equation}\label{eq:meanoverlap}
 Q_1T_{\rm geo}=Q_0\#Q_1,
 \qquad
 R_{\rm geo}=\polar\bigl(U_1^\top(Q_0\#Q_1)U_0\bigr).
\end{equation}
The ambient map is the exact propagator of $\dot v=-\tfrac12Q^{-1}\dot Qv$ along the affine invariant geodesic from $Q_0$ to $Q_1$. The coefficient update is reversible whenever its overlap is nonsingular, so swapping the endpoints gives $R_{\rm geo}^{-1}=R_{\rm geo}^\top$.
\end{proposition}

\begin{proof}
Using $Q_1=Q_0^{1/2}GQ_0^{1/2}$ in \cref{eq:geodesicmap} gives
\begin{align*}
 Q_1T_{\rm geo}
 &=Q_0^{1/2}GQ_0^{1/2}Q_0^{-1/2}G^{-1/2}Q_0^{1/2}\\
 &=Q_0^{1/2}G^{1/2}Q_0^{1/2}=Q_0\#Q_1.
\end{align*}
Substitution in the coefficient overlap proves \cref{eq:meanoverlap}. Along the metric geodesic $Q(s)=Q_0^{1/2}G^sQ_0^{1/2}$, with $0\leq s\leq1$, the selected connection has the constant coefficient
\[
 A^{\min}(s)
 =\tfrac12Q(s)^{-1}\frac{dQ(s)}{ds}
 =\tfrac12Q_0^{-1/2}(\log G)Q_0^{1/2}.
\]
Its propagator is therefore
\[
 T(s)=Q_0^{-1/2}G^{-s/2}Q_0^{1/2},
 \qquad \dot T(s)=-A^{\min}(s)T(s),
 \qquad T(0)=I_n.
\]
At $s=1$, this is the endpoint map in \cref{eq:geodesicmap}. To check reversibility, denote the forward and backward overlaps by $M_{10}$ and $M_{01}$. Symmetry of the geometric mean gives
\[
 M_{01}=U_0^\top(Q_1\#Q_0)U_1
       =\bigl(U_1^\top(Q_0\#Q_1)U_0\bigr)^\top=M_{10}^\top.
\]
For nonsingular overlaps, $\polar(M_{10}^\top)=\polar(M_{10})^\top$, so the backward update is the transpose, and hence the inverse, of the forward update.
\end{proof}

The proposition identifies a standard geometric mean transformation rather than a new matrix mean or a new affine invariant geometry. Its role is to provide the appropriate ambient alignment before the usual polar projection onto the next subspace.

\begin{proposition}[Covariance under constant coordinate changes]\label{prop:coordinate}
Let $C$ be a constant nonsingular change of asset coordinates, with transformed metrics $Q_i'=C^\top Q_iC$ and frames $U_i'=C^{-1}U_i$. The geodesic endpoint map satisfies
\begin{equation}\label{eq:coordinatecovariance}
 T_{\rm geo}'=C^{-1}T_{\rm geo}C,
 \qquad
 U_1'^\top Q_1'T_{\rm geo}'U_0'=U_1^\top Q_1T_{\rm geo}U_0.
\end{equation}
Consequently, the polar coefficient update and the transported asset space vector are independent of a constant reparameterization of the asset coordinates. The same statement holds for risk weighted Procrustes.
\end{proposition}

\begin{proof}
Set $Q_0'=C^\top Q_0C$ and define $O=Q_0^{1/2}C(Q_0')^{-1/2}$. Then
\[
 O^\top O=(Q_0')^{-1/2}C^\top Q_0C(Q_0')^{-1/2}=I_n.
\]
The whitened endpoint matrices and their inverse square roots satisfy
\begin{align*}
 G'&=(Q_0')^{-1/2}C^\top Q_1C(Q_0')^{-1/2}=O^\top GO,\\
 (G')^{-1/2}&=O^\top G^{-1/2}O.
\end{align*}
Using $O^\top=O^{-1}$ on the left and the defining expression for $O$ on the right gives
\begin{align*}
 T_{\rm geo}'
 &=(Q_0')^{-1/2}O^\top G^{-1/2}O(Q_0')^{1/2}\\
 &=C^{-1}Q_0^{-1/2}G^{-1/2}Q_0^{1/2}C
 =C^{-1}T_{\rm geo}C.
\end{align*}
The coefficient overlap is consequently unchanged:
\[
 U_1'^\top Q_1'T_{\rm geo}'U_0'
 =U_1^\top C^{-\top}(C^\top Q_1C)
   (C^{-1}T_{\rm geo}C)C^{-1}U_0
 =U_1^\top Q_1T_{\rm geo}U_0.
\]
For risk weighted Procrustes, the normalization matrix is also unchanged, since
\[
 U_0'^\top Q_1'U_0'=U_0^\top Q_1U_0,
 \qquad \widetilde U_0'=C^{-1}\widetilde U_0.
\]
Thus $U_1'^\top Q_1'\widetilde U_0'=U_1^\top Q_1\widetilde U_0$, proving the same covariance statement for that update.
\end{proof}

\begin{theorem}[Leading error and global transport orders]\label{thm:orders}
Suppose $Q$ and $U$ are three times continuously differentiable on $[0,T]$, their derivatives through order three are bounded, and
\begin{equation}\label{eq:uniformorderassumptions}
 \alpha I_n\preceq Q(t)\preceq\beta I_n,
 \qquad U(t)^\top Q(t)U(t)=I_p,
 \qquad 0<\alpha\leq\beta.
\end{equation}
Write $P=UU^\top Q$, and define the normal metric variation, normal covariant motion, and skew error coefficient by
\begin{align}
 B&=(I_n-P)Q^{-1}\dot Q U,
 &N&=(I_n-P)(\dot U+A^{\min}U),\label{eq:normaldefect}\\
 \mathcal D&=\tfrac14(N^\top QB-B^\top QN),
 &\mathcal D^\top&=-\mathcal D.\label{eq:defectcoefficient}
\end{align}
Let $\Psi(t+h,t)$ be the exact coefficient propagator of \cref{eq:parallel}. For sufficiently small $h$, both endpoint overlaps are nonsingular, uniformly in $t$, and
\begin{align}
 R_{\rm RP}(t+h,t)-\Psi(t+h,t)&=h^2\mathcal D(t)+O(h^3),\label{eq:rpexactdefect}\\
 R_{\rm geo}(t+h,t)-\Psi(t+h,t)&=O(h^3),\label{eq:geolocalorder}\\
 R_{\rm RP}(t+h,t)-R_{\rm geo}(t+h,t)&=h^2\mathcal D(t)+O(h^3).\label{eq:methoddifference}
\end{align}
All remainders are uniform. On any partition with maximal step $h_{\max}$, composition gives global errors $O(h_{\max})$ for risk weighted Procrustes and $O(h_{\max}^2)$ for geodesic Procrustes. If $\mathcal D$ vanishes along the path, risk weighted Procrustes also has global error $O(h_{\max}^2)$.
\end{theorem}

The proof in \cref{app:orders} provides the coefficient in \cref{eq:rpexactdefect}, not merely an upper bound on its order. The terms $B$ and $N$ transform covariantly under a constant change of asset coordinates and by right multiplication under a frame rotation, so $\mathcal D$ transforms by orthogonal conjugation. Its vanishing is therefore independent of these coordinate choices. It vanishes for a constant metric, for a fixed subspace, and whenever $B=0$, including scalar rescalings of a fixed metric. For a fixed subspace, $(I-P)\dot U=0$ and hence $N=B/2$, which proves the assertion directly. At full ambient rank, both updates coincide exactly, as follows by whitening the old metric and taking the old frame to be the identity. At rank one their coefficient maps equal the exact local transport on a sufficiently fine grid, since the only continuous orthogonal coefficient evolution is the identity. These cases distinguish the present result from the established second order accuracy of polar transport under a fixed metric \citep{littlejohn2022,harland2023,zhang2026}.

\begin{proposition}[Sharpness with a commuting metric]\label{prop:sharp}
In $\R^3$, let $e_1,e_2,e_3$ be the standard basis, set $a=(e_1+e_3)/\sqrt2$, $b=(e_1-e_3)/\sqrt2$, and define
\begin{equation}\label{eq:sharppath}
 L=be_2^\top-e_2b^\top,
 \quad Q(t)=\diag(e^t,1,e^{-t}),
 \quad U(t)=Q(t)^{-1/2}e^{tL}[\,a\ \ e_2\,].
\end{equation}
For $0<T<\pi/2$, the metric commutes with its derivative, the exact coefficient propagator is $I_2$, and every sufficiently small geodesic Procrustes step is exactly $I_2$. On a uniform partition with step $h=T/n$, the risk weighted Procrustes product satisfies
\begin{equation}\label{eq:sharpglobal}
 R_{{\rm RP},n-1}\cdots R_{{\rm RP},0}
 =I_2+\frac{h\sin T}{4}
 \begin{pmatrix}0&-1\\1&0\end{pmatrix}+O(h^2).
\end{equation}
Thus its first order global error is sharp even within the class of commuting metric paths.
\end{proposition}

\begin{proof}
\proofstep{Exact continuous and geodesic transport.}
Write $Z(t)=e^{tL}[\,a\ \ e_2\,]$ and $b_t=e^{tL}b$. Since $La=0$, $Le_2=b$, and $Lb=-e_2$, these vectors have the explicit expressions
\[
 Z(t)=[\,a\ \ z_t\,],
 \qquad z_t=\cos t\,e_2+\sin t\,b,
 \qquad b_t=\cos t\,b-\sin t\,e_2.
\]
The metric commutes with its derivative, so $\mathcal B=0$ in \cref{eq:whiteningconnection}. Moreover,
\[
 \dot Z(t)=[\,0\ \ b_t\,],
 \qquad Z(t)^\top\dot Z(t)=0,
\]
which gives $\omega(t)=0$ and hence $\Psi(t+h,t)=I_2$. The geodesic overlap reduces to
\[
 Z(t+h)^\top Z(t)
 =\begin{pmatrix}1&0\\0&\cos h\end{pmatrix}.
\]
For sufficiently small $h$, this matrix is positive definite and its polar factor is exactly $I_2$.

\proofstep{Nonzero local defect.}
In whitened coordinates, the two normal terms in \cref{eq:normaldefect} are
\[
 Q^{1/2}N=b_t\begin{pmatrix}0&1\end{pmatrix},
 \qquad
 Q^{1/2}B=b_t\begin{pmatrix}\cos t&0\end{pmatrix}.
\]
Since $b_t^\top b_t=1$, substitution in \cref{eq:defectcoefficient} gives
\[
 \mathcal D(t)=\frac{\cos t}{4}
 \begin{pmatrix}0&-1\\1&0\end{pmatrix}.
\]
Thus \cref{eq:rpexactdefect} yields $R_{{\rm RP},k}=I_2+h^2\mathcal D(t_k)+O(h^3)$ on the uniform grid $t_k=kh$.

\proofstep{Global accumulation.}
The local remainders sum to $O(nh^3)=O(h^2)$, and products of two or more local defects contribute $O(n^2h^4)=O(h^2)$. Consequently,
\[
 R_{{\rm RP},n-1}\cdots R_{{\rm RP},0}
 =I_2+h^2\sum_{k=0}^{n-1}\mathcal D(t_k)+O(h^2).
\]
The Riemann sum satisfies
\[
 h^2\sum_{k=0}^{n-1}\mathcal D(t_k)
 =h\int_0^T\mathcal D(t)\,dt+O(h^2)
 =\frac{h\sin T}{4}\begin{pmatrix}0&-1\\1&0\end{pmatrix}+O(h^2).
\]
This proves \cref{eq:sharpglobal}, whose leading coefficient is nonzero for $0<T<\pi/2$.
\end{proof}

The example separates two mechanisms that should not be conflated. Noncommutativity of the metric determines the ambient discrepancy of principal square root whitening, whereas the order loss of risk weighted Procrustes is determined by the coupling in \cref{eq:defectcoefficient}. Commutation removes the former but need not remove the latter.

\begin{proposition}[Symmetric Procrustes alternative]\label{prop:symmetric}
Under the hypotheses of \cref{thm:orders}, let $F=R_{\rm RP}(t_1,t_0)$ and $B=R_{\rm RP}(t_0,t_1)$ denote the forward and backward risk weighted updates. For sufficiently close endpoints, the map
\begin{equation}\label{eq:symmetricprocrustes}
 R_{\rm sym}(t_1,t_0)=\polar\left(\frac{F+B^\top}{2}\right)
\end{equation}
is well defined, reversible, and covariant under changes of endpoint frames and constant asset coordinates. It has local error $O(h^3)$ and global error $O(h_{\max}^2)$ for the selected connection.
\end{proposition}

\begin{proof}
\proofstep{Cancellation of the local defect.}
Write $\Psi=\Psi(t+h,t)$. In the same smooth frame, the forward and backward expansions from \cref{eq:rpexactdefect} give
\begin{align*}
 F&=\Psi+h^2\mathcal D(t)+O(h^3),\\
 B^\top&=\Psi-h^2\mathcal D(t+h)+O(h^3).
\end{align*}
Because $\mathcal D$ is continuously differentiable, their average satisfies
\[
 \frac{F+B^\top}{2}
 =\Psi+\frac{h^2}{2}\bigl(\mathcal D(t)-\mathcal D(t+h)\bigr)+O(h^3)
 =\Psi+O(h^3).
\]
This matrix is nonsingular for small $h$. Smoothness of the polar map near the orthogonal matrix $\Psi$, together with $\polar(\Psi)=\Psi$, therefore gives
\[
 R_{\rm sym}(t+h,t)=\Psi(t+h,t)+O(h^3).
\]

\proofstep{Reversibility and covariance.}
Exchanging the endpoints transposes the average, so
\[
 R_{\rm sym}(t,t+h)
 =\polar\!\left(\frac{B+F^\top}{2}\right)
 =R_{\rm sym}(t+h,t)^\top.
\]
For endpoint frame rotations $U_i'=U_iO_i$, both terms transform as
\[
 F'=O_1^\top FO_0,
 \qquad (B')^\top=O_1^\top B^\top O_0.
\]
Orthogonal covariance of the polar factor proves the frame transformation law, while constant asset coordinate covariance follows from \cref{prop:coordinate}. Finally, all updates are orthogonal, so the accumulation argument in \cref{app:orders} gives a global bound proportional to
\[
 \sum_k h_k^3\leq T h_{\max}^2,
\]
which establishes the stated second order convergence.
\end{proof}

Symmetrization is a standard device in geometric integration \citep{hairer2006}, and the proposition applies it to the explicit error in \cref{eq:defectcoefficient}. It is included to make clear that the geometric mean rule is not the only second order endpoint implementation. Given dense metrics and supplied frames, the additional work for \cref{eq:symmetricprocrustes} is $O(n^2p+np^2+p^3)$, involving products with the metrics and matrix functions of dimension $p$. It avoids the matrix functions of dimension $n$ in a direct evaluation of the geometric mean. The two second order rules need not agree at a finite step, so numerical and financial comparisons include both.

\subsection{Stability under estimated geometry}

The continuous theory is meaningful for estimation only if small errors in frames and connections produce controlled portfolio errors. The following result isolates the required quantities.

\begin{assumption}[Uniform regularity]\label{ass:regularity}
On $[0,T]$, the true and estimated metrics are bounded between $\alpha I_n$ and $\beta I_n$ for constants $0<\alpha\leq\beta$. The true and estimated frames have been placed in a common orthogonal gauge, are continuously differentiable, and have skew-symmetric connection forms in that gauge.
\end{assumption}

\begin{theorem}[Stability of transported portfolios]\label{thm:stability}
Let $c$ and $\widehat c$ solve the true and estimated parallel equations, and set $v=Uc$ and $\widehat v=\widehat U\widehat c$. Under \ref{ass:regularity}, define
\begin{equation}\label{eq:errors}
 \varepsilon_U(t)=\norm[2]{\widehat U_t-U_t},
 \qquad
 \varepsilon_\omega(t)=\norm[2]{\widehat\omega_t-\omega_t}.
\end{equation}
Then
\begin{equation}\label{eq:stability}
 \norm[Q_t]{\widehat v_t-v_t}
 \leq
 \sqrt\beta\,\varepsilon_U(t)\norm{\widehat c_0}
 +\norm{\widehat c_0-c_0}
 +\norm{c_0}\int_0^t\varepsilon_\omega(s)\,ds.
\end{equation}
\end{theorem}

\begin{proof}
\proofstep{Coefficient error.}
Let $e_t=\widehat c_t-c_t$ and let $\widehat\Psi(t,s)$ be the propagator generated by $-\widehat\omega$. Subtracting the two parallel equations gives
\[
 \dot e_t=-\widehat\omega_te_t-(\widehat\omega_t-\omega_t)c_t.
\]
Variation of constants therefore yields
\[
 e_t=\widehat\Psi(t,0)e_0
     -\int_0^t\widehat\Psi(t,s)(\widehat\omega_s-\omega_s)c_s\,ds.
\]
Since $\widehat\Psi(t,s)$ is orthogonal and $\|c_s\|=\|c_0\|$, it follows that
\[
 \|e_t\|
 \leq\|e_0\|+\|c_0\|\int_0^t\varepsilon_\omega(s)\,ds.
\]

\proofstep{Asset space error.}
Separate the frame error from the coefficient error through
\[
 \widehat v_t-v_t
 =(\widehat U_t-U_t)\widehat c_t+U_te_t.
\]
Using $Q_t\preceq\beta I_n$, $U_t^\top Q_tU_t=I_p$, and norm preservation of $\widehat c_t$ gives
\begin{align*}
 \|\widehat v_t-v_t\|_{Q_t}
 &\leq\sqrt\beta\,\|\widehat U_t-U_t\|_2\,\|\widehat c_t\|+\|e_t\|\\
 &\leq\sqrt\beta\,\varepsilon_U(t)\|\widehat c_0\|
       +\|\widehat c_0-c_0\|
       +\|c_0\|\int_0^t\varepsilon_\omega(s)\,ds,
\end{align*}
which is \cref{eq:stability}.
\end{proof}

\begin{corollary}[Spectral subspace control]\label{cor:spectral}
Let $Z=Q^{1/2}U$ and $\widehat Z=\widehat Q^{1/2}\widehat U$. Suppose their column spaces are spectral clusters of symmetric whitened operators $L$ and $\widehat L$. Assume that the cluster of $L$ is separated from the remaining spectrum by $\delta>0$, that the cluster of $\widehat L$ is its corresponding perturbation, and that
\begin{equation}\label{eq:gapconditionmain}
 \norm[2]{\widehat L-L}<\delta/2,
\end{equation}
then an orthogonal alignment exists for which
\begin{equation}\label{eq:daviskahan}
 \norm[2]{\widehat U-U}
 \leq
 \frac{2\sqrt2}{\sqrt\alpha\,\delta}\norm[2]{\widehat L-L}
 +\frac{1}{2\alpha^{3/2}}\norm[2]{\widehat Q-Q}.
\end{equation}
\end{corollary}

\begin{proof}
The Davis--Kahan theorem and orthogonal Procrustes alignment provide an orthogonal matrix $R$ for which
\begin{align*}
 \|\widehat ZR-Z\|_2
 &\leq\sqrt2\,\|\widehat Z\widehat Z^\top-ZZ^\top\|_2\\
 &\leq\frac{2\sqrt2}{\delta}\,\|\widehat L-L\|_2.
\end{align*}
To return to asset coordinates, separate the subspace error from the metric error:
\[
 \widehat UR-U
 =\widehat Q^{-1/2}(\widehat ZR-Z)
  +(\widehat Q^{-1/2}-Q^{-1/2})Z.
\]
Since $\|Z\|_2=1$ and $\|\widehat Q^{-1/2}\|_2\leq\alpha^{-1/2}$, this gives
\[
 \|\widehat UR-U\|_2
 \leq\frac{2\sqrt2}{\sqrt\alpha\,\delta}\,\|\widehat L-L\|_2
      +\|\widehat Q^{-1/2}-Q^{-1/2}\|_2.
\]
The operator Lipschitz bound for the inverse square root on $[\alpha,\infty)$ is
\[
 \|\widehat Q^{-1/2}-Q^{-1/2}\|_2
 \leq\frac{1}{2\alpha^{3/2}}\,\|\widehat Q-Q\|_2.
\]
Combining these inequalities proves \cref{eq:daviskahan} in the aligned frame \citep{daviskahan1970,higham2008}.
\end{proof}

\begin{corollary}[Turnover perturbation]\label{cor:turnover}
On a decision grid, let $e_k=\widehat\pi_{t_k}-\pi_{t_k}$ and $e_k^-=\widehat\pi_{t_k^-}-\pi_{t_k^-}$. Then
\begin{equation}\label{eq:turnbound}
 |\operatorname{TO}(\widehat\pi)-\operatorname{TO}(\pi)|
 \leq \frac12\sum_{k=1}^{K}(\norm[1]{e_k}+\norm[1]{e_k^-}).
\end{equation}
\end{corollary}

\begin{proof}
At each decision date, set $\Delta_k=\pi_{t_k}-\pi_{t_k^-}$ and $\widehat\Delta_k=\widehat\pi_{t_k}-\widehat\pi_{t_k^-}$. The reverse triangle inequality gives
\begin{align*}
 \bigl|\|\widehat\Delta_k\|_1-\|\Delta_k\|_1\bigr|
 &\leq\|\widehat\Delta_k-\Delta_k\|_1\\
 &=\|e_k-e_k^-\|_1
 \leq\|e_k\|_1+\|e_k^-\|_1.
\end{align*}
Summing over decisions and multiplying by one half proves \cref{eq:turnbound}.
\end{proof}

The pretrade term cannot in general be replaced by the preceding target error: it also contains the realized drift of holdings. Under uniformly positive portfolio gross returns and bounded asset returns, the buy and hold update is locally Lipschitz, so $e_k^-$ is controlled recursively by the preceding posttrade error and the intervening return path. Connection error depends on temporal derivatives of the estimated metric and frame in addition to their pointwise estimation errors, as quantified in the appendix. Degeneration of the spectral gap or an endpoint overlap compromises the corresponding frame or transport stability bound, so the resulting dates require separate treatment from the smooth paths of fixed rank considered here.

\section{Numerical Results}

\subsection{Experimental design and comparison set}

The experiments compare independent local refitting, ambient projection, Euclidean Procrustes, risk weighted Procrustes, symmetric Procrustes, and the geometric mean update, separating the effects of target persistence from those of the transport rule. The manufactured comparison also includes principal square root whitening to assess its consistency with the selected connection alongside its preservation of endpoint inner products.

The manufactured calculations use double precision, while every market comparison supplies the methods with the same momentum signal computed from preceding observations, Ledoit--Wolf covariance estimate, investable subspace, equal weight feasible anchor, active risk budget, gross exposure cap, and proportional costs. The daily panels use a 504 day covariance window and rebalance every 21 trading days, with representation ranks of five for the 48 industry panel and eight for the 100 portfolio panel. Costs are 10 and 15 basis points. The primary persistence update assigns 25 percent weight to the new local target. The reported sensitivity grid assigns 10, 25, 50, or 75 percent to the newly estimated target. Weekly and quarterly rebalancing are reported separately as frequency sensitivity. Between decisions the portfolios are buy and hold. At each decision, turnover is computed against the drifted pretrade portfolio and the proportional cost is deducted multiplicatively from wealth before the next holding period return.

\subsection{Connection, gauge, and motion verification}

The first manufactured problem uses four asset coordinates, a rank two moving subspace, and a nonconstant metric with an analytic coefficient solution. A fourth order Runge--Kutta solver supplies an independent numerical reference for the continuous connection against which integration and gauge covariance are checked.

\begin{table}[ht]
\centering
\caption{Manufactured connection and gauge checks.}\label{tab:gate1}
\begin{tabularx}{0.92\textwidth}{@{}Yl@{}}
\toprule
Quantity & Result\\
\midrule
Observed reference integration order & 4.000 on the finest refinement\\
Maximum random gauge transport error, 100 gauges & $4.58\times10^{-10}$\\
Maximum projector gauge error & $5.92\times10^{-16}$\\
Maximum connection covariance error & $2.97\times10^{-15}$\\
Maximum connection skew error & $5.44\times10^{-15}$\\
Maximum motion decomposition residual & $3.15\times10^{-16}$\\
Maximum normal orthogonality residual & $3.61\times10^{-16}$\\
Geometric rebalancing energy & $9.49\times10^{-2}$\\
\bottomrule
\end{tabularx}
\end{table}

The diagnostics in \cref{tab:gate1} satisfy the specified tolerances, including the reference integration and gauge checks. The nonuniqueness counterexample uses a constant identity metric and compares the zero connection with a nonzero skew coefficient. Although both are compatible, their terminal transports differ by $0.686$, whereas the minimum norm objective selects the zero coefficient.

\subsection{Noncommuting metric path and convergence order}

The second manufactured problem uses six assets and a rank three representation. The metric eigenspaces rotate while unequal eigenvalues vary, so the metric and its derivative do not commute. A diagonal path provides the commuting control. The continuous coefficient equation is solved with DOP853 at relative tolerance $2\times10^{-12}$.

\begin{table}[ht]
\centering
\caption{Global convergence to the selected induced connection.}\label{tab:orders}
\begin{tabularx}{0.92\textwidth}{@{}Yrrr@{}}
\toprule
Method & Noncommuting order & Finest risk norm error & Commuting order\\
\midrule
Risk weighted Procrustes & 1.001 & $2.56\times10^{-4}$ & 0.991\\
Principal square root map & $-0.000$ & $6.36\times10^{-2}$ & 2.000\\
Geometric mean update & 2.000 & $3.47\times10^{-7}$ & 2.000\\
Symmetric Procrustes & 2.000 & $2.75\times10^{-7}$ & 2.000\\
\bottomrule
\end{tabularx}
\end{table}

\begin{figure}[ht]
\centering
\includegraphics[width=\textwidth]{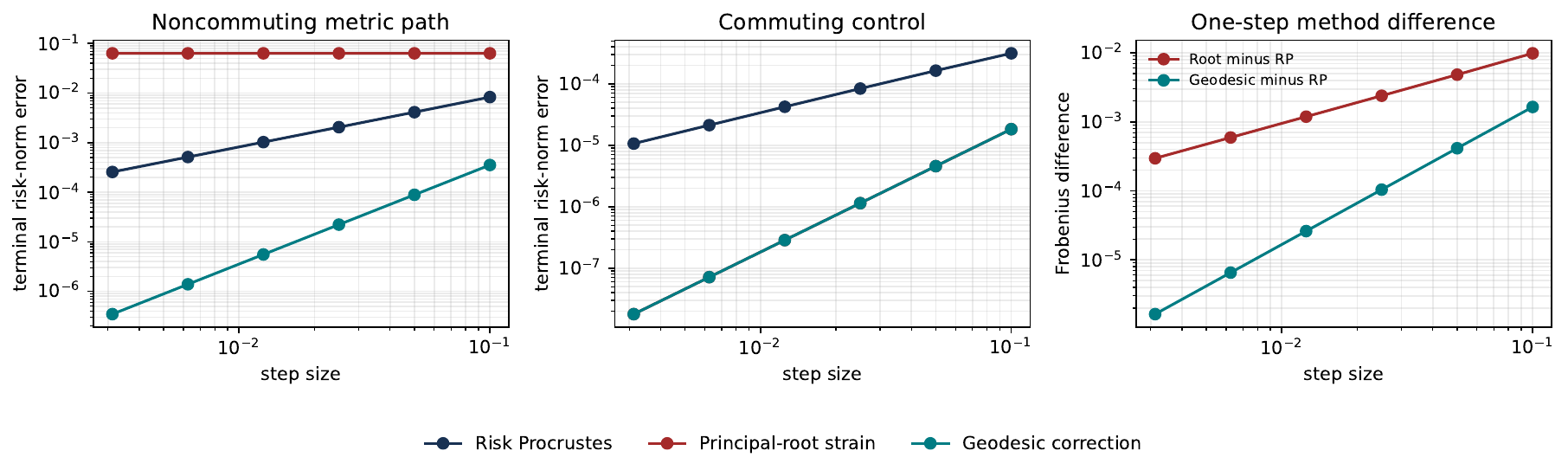}
\caption{Endpoint convergence comparison. Principal square root map converges to a different connection on the noncommuting path. Risk Procrustes is first order and the geometric mean update is second order.}\label{fig:orders}
\end{figure}

The principal root error plateaus under refinement in the noncommuting case, exactly as \cref{thm:obstruction} predicts. Its single step difference from risk Procrustes is first order. The geodesic and risk Procrustes updates differ at second order, while the geometric mean method converges globally at order two. The projected discrepancy norm is $9.48\times10^{-2}$ and the metric derivative commutator norm is $4.54$, whereas both quantities vanish in the commuting control. Every endpoint map passes its isometry test, confirming the distinction between endpoint norm preservation and consistency with the intended connection. Maximum gauge error across all methods and both paths is $2.41\times10^{-14}$. Symmetric Procrustes also converges at order two, with a smaller error than the geometric mean rule on the noncommuting test and a larger error on the commuting test. Its reversal, gauge, and constant coordinate residuals are below $2\times10^{-15}$. These comparisons demonstrate equal asymptotic order without a uniform finite step ranking. A separate nonsingular nonorthogonal change of asset coordinates gives maximum covariance error $4.27\times10^{-15}$ for the geodesic map and the two metric based coefficient updates, numerically confirming \cref{prop:coordinate}. Figure~\ref{fig:orders} displays the local and global refinement results.

\subsection{Exact sharpness and the local error coefficient}

The rank two construction in \cref{prop:sharp} provides a reference that requires no differential equation solver. We set $T=1/2$, for which the exact coefficient map is $I_2$ and the predicted operator norm error divided by the mesh size converges to $\sin(1/2)/4$. Table~\ref{tab:sharp} reports the refinement. The fitted order on the three finest grids is $1.000$, while the geometric mean update remains at machine precision. This example verifies both sharpness and the distinction between metric commutativity and the normal coupling that controls the Procrustes error.

\begin{table}[ht]
\centering
\caption{Analytic commuting example. The reference is exactly $I_2$, not a refined computation of either numerical method. Errors are matrix operator norms.}\label{tab:sharp}
\begin{tabular}{@{}rrrr@{}}
\toprule
Steps & $\|E_{\rm RP}\|_2$ & $10^3\|E_{\rm RP}\|_2/h$ & $\|E_{\rm geo}\|_2$\\
\midrule
32 & $1.88\times10^{-3}$ & $120.040$ & $1.23\times10^{-16}$\\
64 & $9.37\times10^{-4}$ & $119.949$ & $7.96\times10^{-16}$\\
128 & $4.68\times10^{-4}$ & $119.903$ & $2.29\times10^{-16}$\\
256 & $2.34\times10^{-4}$ & $119.880$ & $2.85\times10^{-16}$\\
\bottomrule
\end{tabular}
\end{table}

A separate local calculation on the six dimensional noncommuting path tests the coefficient itself. At time $0.37$, the norm of $\mathcal D$ is $0.118$. The discrepancy between $(R_{\rm RP}-R_{\rm geo})/h^2$ and $\mathcal D$ decreases from $2.374\times10^{-4}$ at step $0.005$ to $1.185\times10^{-4}$ at step $0.0025$, as the linear remainder in \cref{eq:methoddifference} predicts. The geometric mean identity, reversal law, whitening identity, and coincidence at full ambient rank have residuals below $5\times10^{-15}$. The derivative of the inverse square root is evaluated through its continuous divided difference formula, including repeated eigenvalues, rather than by subtracting nearly equal eigenvalues.

\subsection{Historical geometry stress test}

The theoretical distinction between first and second order transport should matter most when the defect coefficient in \cref{eq:defectcoefficient} is large. We therefore add a historical geometry diagnostic using the 48 industry panel and the same Ledoit--Wolf metric and rank five representation used in the portfolio experiment. The diagnostic does not use portfolio returns, transaction costs, persistence parameters, or certainty equivalent. At each five day state, $\dot Q$ and $\dot U$ are estimated from the preceding five observations after a gauge only alignment of the risk whitened frames. These estimates give an ex ante value $\widehat{\mathcal D}_t$. We then compare it with the separation of the risk weighted Procrustes and geometric mean coefficient maps over the next five observations. The scale predicted by \cref{eq:methoddifference} is
\[
 \frac{\|R_{{\rm RP},t+5}-R_{{\rm geo},t+5}\|_2}{5^2}.
\]
The same comparison is repeated with symmetric Procrustes as an independently constructed second order endpoint rule.

Across 1,727 historical comparisons, the Spearman correlation between $\|\widehat{\mathcal D}_t\|_2$ and the subsequent scaled risk Procrustes versus geometric mean separation is $0.458$, with a moving block bootstrap 95 percent interval of $[0.394,0.518]$. The mean subsequent separation rises monotonically across ex ante defect quartiles from $1.57\times10^{-5}$ in the lowest quartile to $8.21\times10^{-5}$ in the highest. The top to bottom ratio is $5.24$, and the paired quartile difference has a moving block bootstrap interval of $[4.04\times10^{-5},1.01\times10^{-4}]$. The median ratio between the observed scaled separation and the predicted defect norm is $0.977$, with a bootstrap interval of $[0.948,1.020]$.

\begin{figure}[ht]
\centering
\includegraphics[width=0.98\textwidth]{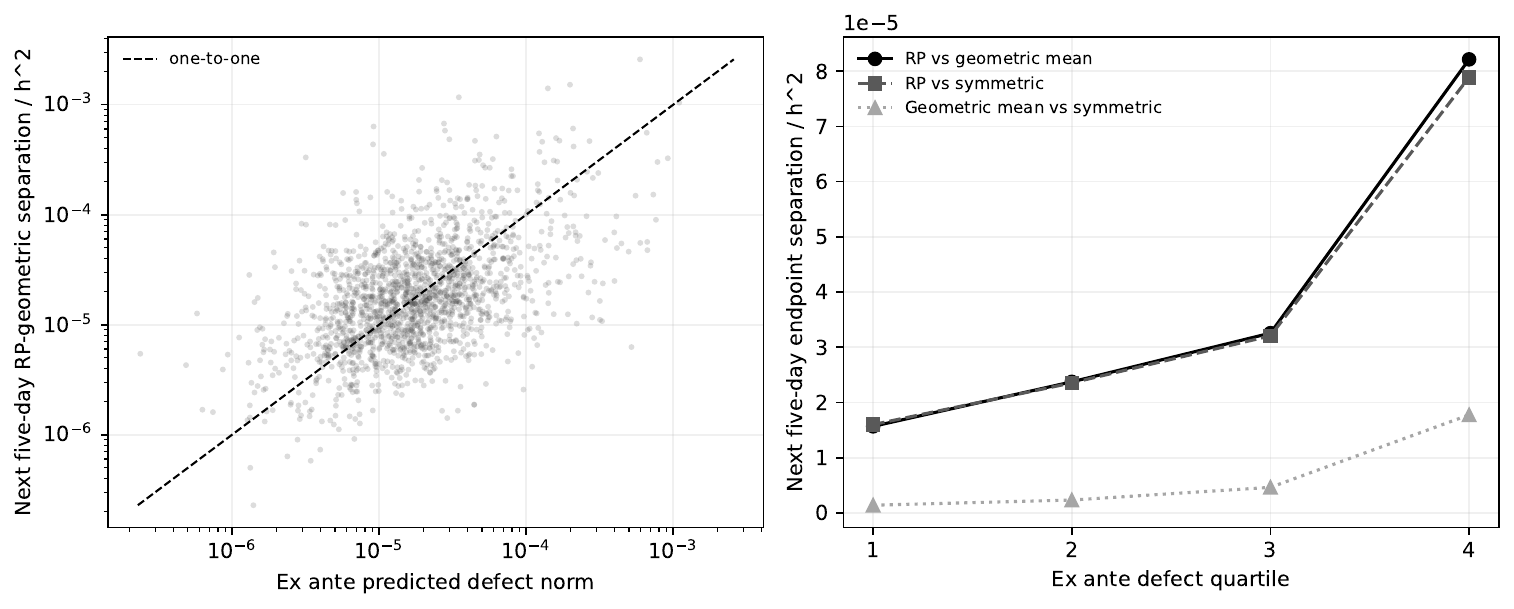}
\caption{Historical geometry stress diagnostic for the 48 industry panel. Left: the defect norm estimated from the preceding five observations against the scaled separation of risk weighted Procrustes and geometric mean transport over the next five observations. The dashed line is equality. Right: mean next interval endpoint separation by ex ante defect quartile. The two second order rules remain substantially closer to each other than either is to risk weighted Procrustes.}\label{fig:historicaldefect}
\end{figure}

The symmetric construction provides a separate check that this pattern is not specific to the geometric mean rule. Its separation from risk weighted Procrustes has correlation $0.458$ with the same ex ante defect and a top to bottom quartile ratio of $4.94$. By contrast, the median scaled separation between the geometric mean and symmetric second order updates is $7.62\times10^{-7}$, compared with $1.65\times10^{-5}$ between risk weighted Procrustes and geometric mean transport. Thus the two second order constructions are about $21.7$ times closer to each other in median scaled endpoint distance than risk weighted Procrustes is to the geometric mean update. Figure~\ref{fig:historicaldefect} reports the pointwise and quartile diagnostics.

\subsection{Historical portfolio implementation}

The financial exercise uses daily value weighted returns from the Kenneth R. French Data Library \citep{famafrench1993,frenchdata2026}. The first panel contains 48 industry portfolios and the second contains 100 portfolios formed at the intersections of ten size groups and ten book to market groups. Source returns are percentages and are converted to decimal units. Missing value codes are treated as unavailable observations rather than as returns, and the source trading calendar is preserved when forming estimation and holding windows.

The candidate sample is 2 January 1990 through 30 April 2026. For each panel, we select the longest uninterrupted run in which all portfolio returns are available, measured on the original source calendar, with the earliest run selected in a length tie. This rule depends on availability rather than strategy performance and is identical for all methods. It gives 9,149 observations from 2 January 1990 through 30 April 2026 for the industry panel and 3,774 observations from 1 July 2009 through 28 June 2024 for the 100 portfolio panel. Dates separated by missing source rows are never treated as consecutive trading observations. After the initial 504 observations, the primary tests contain 8,645 returns and 412 decisions, and 3,270 returns and 156 decisions, respectively. The panels therefore cover different periods, and comparisons are made within each panel rather than between their unconditional performance levels.

Signals and covariance estimates use observations preceding each decision date, while sample selection and the source files are retrospective. The comparison is therefore conditional on the specified research panels, with historical data revisions outside the information restrictions imposed by the experiment. Signals, constraints, costs, and persistence settings are common to all methods, as specified in \cref{app:protocol}.

\begin{table}[ht]
\centering
\caption{Primary results after proportional costs. Certainty equivalent (CE) uses arithmetic annualized mean minus one half of risk aversion three times annualized variance. Turnover (TO) is annualized over the actual number of evaluated daily observations. The last column is geometric mean transport minus local refitting, with a paired moving block bootstrap interval.}\label{tab:markets}
\small
\begin{tabular}{@{}lrrrrr@{}}
\toprule
Universe & CE local & CE geo. & TO local & TO geo. & $\Delta$CE, bp [95\%]\\
\midrule
48 industries & 8.458\% & 8.364\% & 2.997 & 1.557 & $-9.37\;[-95.93,81.48]$\\
100 size--value & 6.781\% & 5.609\% & 4.654 & 2.658 & $-117.20\;[-279.61,-1.64]$\\
\bottomrule
\end{tabular}
\end{table}

At the primary persistence weight, geometric mean transport reduces annualized turnover by $48.06\%$ in the industry panel and $42.89\%$ in the 100 portfolio panel relative to independent local refitting, with paired intervals excluding zero in both cases. The corresponding annual certainty equivalent differences are $-9.37$ and $-117.20$ basis points. The industry interval spans zero, whereas the nominal interval for the 100 portfolio panel lies below zero, indicating a performance cost associated with slower adjustment that exceeds the reduction in transaction costs under the specified signal and constraints.

With the persistence parameter held fixed, geometric mean transport changes annual certainty equivalent relative to risk weighted Procrustes by $-1.56$ basis points in the industry panel and $0.26$ basis points in the 100 portfolio panel, with turnover differences of $0.016\%$ and $-0.210\%$ of the corresponding Procrustes levels. Although the industry certainty equivalent interval lies below zero, the estimated difference is small in magnitude, while the 100 portfolio interval spans zero. Relative to symmetric Procrustes, the certainty equivalent differences are $-0.58$ and $-0.76$ basis points, with both intervals spanning zero. At these settings, the effects of changing the transport rule are therefore small compared with those associated with target persistence. The persistence and rebalance frequency grids assess the sensitivity of this comparison to the adjustment horizon, with the primary specification held fixed throughout.

\subsection{Composition diagnostic}

The composition diagnostic compares direct endpoint updates over 21 or 63 trading observations with geometric mean updates composed at five observation intervals, including a shorter final interval where required. Starting dates are separated by 65 observations, so comparison intervals are disjoint at both horizons, and covariance estimates use preceding observations throughout. Because the reference is constructed from geometric mean updates, the diagnostic measures consistency with this specified discrete path rather than error relative to an independently known continuous transport.

\begin{table}[ht]
\centering
\caption{Direct versus composed discrepancy in the industry panel. Method columns report marginal median relative risk norm discrepancies. The final column reports the paired mean difference between the geometric mean and ambient rules, with its bootstrap interval.}\label{tab:path}
\small
\begin{tabular}{@{}rrrrrr@{}}
\toprule
Days & Pairs & Ambient & Risk Proc. & Geometric & Geo.--ambient [95\%]\\
\midrule
21 & 133 & 2.305\% & 0.162\% & 0.131\% & $-3.730\%\;[-4.985,-2.626]$\\
63 & 133 & 6.307\% & 0.803\% & 0.675\% & $-7.555\%\;[-9.448,-5.869]$\\
\bottomrule
\end{tabular}
\end{table}

\begin{figure}[ht]
\centering
\includegraphics[width=0.98\textwidth]{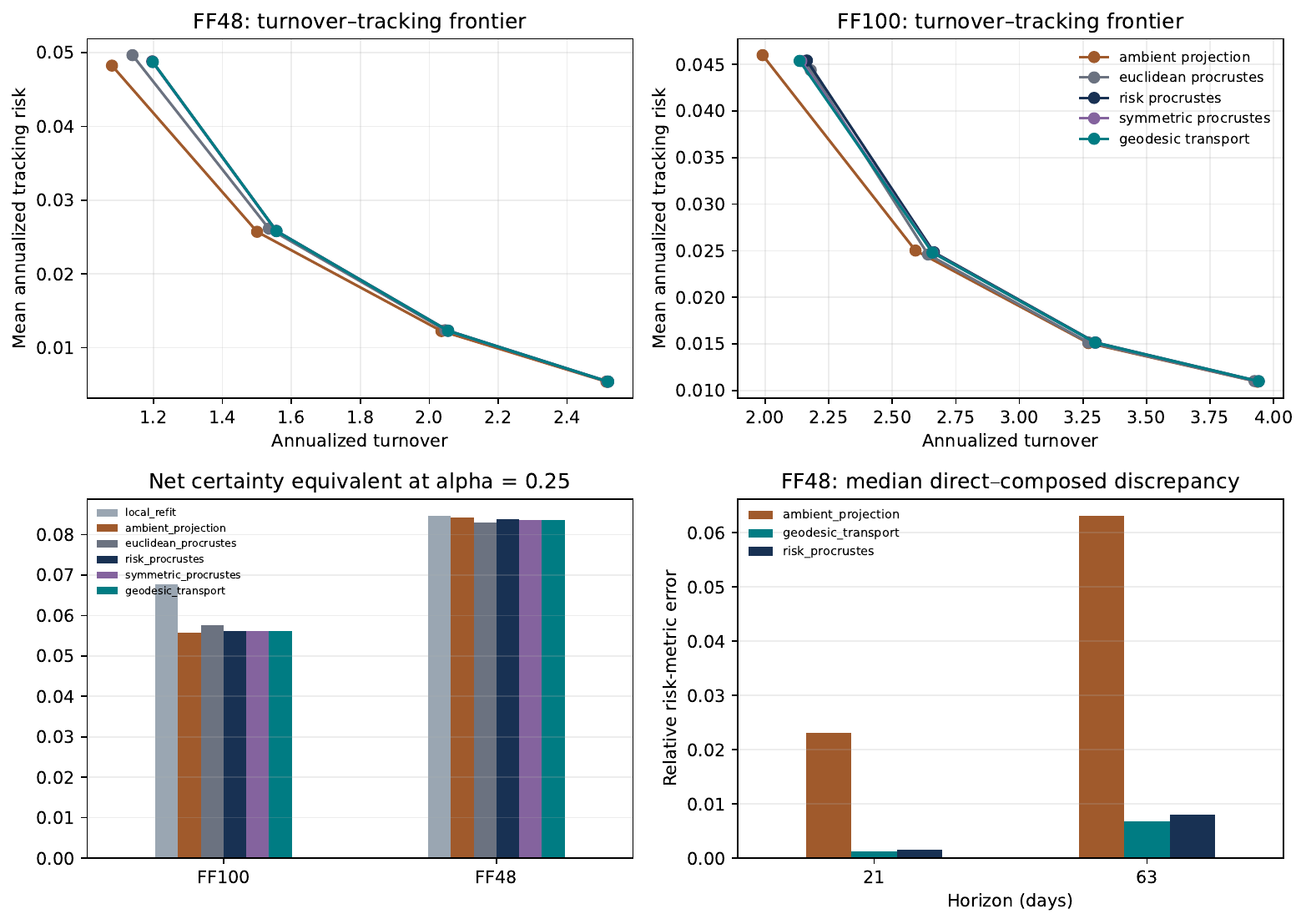}
\caption{Historical implementation comparison. Top: turnover and tracking across the persistence grid. Bottom left: net certainty equivalent at the primary weight. Bottom right: median direct versus composed discrepancy, using the geometric mean rule to construct the reference.}\label{fig:empirical}
\end{figure}

Both metric based updates have smaller discrepancies from the composed reference than ambient projection, with the geometric mean and risk weighted Procrustes results closer to each other. This comparison characterizes agreement with the specified discrete path, while the exact example and manufactured differential equation tests provide the independent references used to assess convergence to continuous transport. Figure~\ref{fig:empirical} presents the portfolio implementation results together with the composition diagnostic.

\subsection{Numerical and economic interpretation}

The manufactured examples recover the predicted local defect, the sharp first order behavior of risk weighted Procrustes, and the second order convergence of the corrected updates. On the noncommuting test path, principal square root whitening converges to a different connection because the projected discrepancy is nonzero, consistently with the criterion in \cref{thm:obstruction}. The historical comparisons isolate a separate implementation effect, with target persistence producing substantial turnover reductions and larger certainty equivalent changes than the choice among the metric based transport rules.

The convergence bounds apply to smooth paths of fixed rank with uniformly positive metrics and nonsingular endpoint overlaps, while the additional spectral gap condition concerns estimation of a selected eigenspace. Their application to rolling financial estimates consequently depends on the regularity of the estimated paths and the derivative errors entering the stability bounds. The portfolio experiments are conditional on the retrospective research panels, momentum signal, constraints, and proportional cost specification, which excludes endogenous market impact. Comparisons are made within each panel because the available sample periods differ, and statistical uncertainty is assessed conditional on that design.

\section{Conclusion}

The analysis identifies the leading local error of risk weighted Procrustes when the risk metric and represented subspace vary simultaneously. This error is the skew coupling between normal covariant motion and normal metric deformation in \cref{eq:defectcoefficient}, and the exact rank two construction shows that it can reduce global accuracy to first order even for a metric that commutes with its derivative. Geometric mean alignment and symmetric Procrustes cancel the term, yielding reversible endpoint schemes with second order accuracy on nonuniform partitions. The symmetric construction achieves this order using matrix operations in the subspace dimension.

These results extend the comparison of induced transport schemes to a varying ambient metric through an explicit defect formula, its sharpness, and its cancellation. The vanishing conditions recover second order Procrustes accuracy in the relevant restricted geometries, while the stability estimates distinguish numerical error from uncertainty in estimated frames and connections. Together, the results specify the geometric conditions governing the choice of endpoint update and the regularity needed to control its accumulated error.

The historical geometry diagnostic links the theoretical defect to estimated market geometry without using portfolio performance: the ex ante defect predicts the subsequent separation of risk weighted Procrustes from both second order updates, while the geometric mean and symmetric constructions remain much closer to each other. The portfolio application addresses a different question. Holding target persistence fixed leaves the economic differences among the metric based transport rules small at the primary settings, while persistence itself materially changes turnover and adjustment lag. The experiments therefore separate numerical transport accuracy from the economic consequences of delaying movement toward a newly estimated target.

\section*{Declarations}

\noindent\textbf{Funding.} The author received no external funding for this research.

\noindent\textbf{Competing interests.} The author is employed by Miralta Finance Bank S.A. The employer had no role in the study design, analysis, interpretation, or decision to submit. The author reports no other competing interests. The views expressed are those of the author and do not necessarily reflect those of Miralta Finance Bank S.A.

\clearpage

\appendix

\section{Proof of the endpoint orders}\label{app:orders}

The proof of \cref{thm:orders} separates the endpoint expansions, the polar projection, the local error coefficient, and its accumulation over a nonuniform partition. Every calculation below uses the same smooth reference frame.

\subsection{Local coordinates and frame constraints}

Apply a constant whitening transformation at the left endpoint, so that $Q(0)=I_n$, and write
\begin{equation}\label{eq:taylorq}
 \begin{aligned}
 Q(h)&=I_n+hH+\tfrac12h^2J+O(h^3),\\
 U(h)&=U+hV+\tfrac12h^2W+O(h^3).
 \end{aligned}
\end{equation}
Here $H$ and $J$ are symmetric. Differentiating $U(h)^\top Q(h)U(h)=I_p$ once and twice yields
\begin{align}
 U^\top V+V^\top U+U^\top HU&=0,\label{eq:constraint1}\\
 \begin{split}
 U^\top W+W^\top U+2V^\top V
 +2V^\top HU\\
 {}+2U^\top HV+U^\top JU
 \end{split}
 &=0.\label{eq:constraint2}
\end{align}
These identities will determine the skew parts of both endpoint expansions.

\subsection{Risk weighted Procrustes expansion}

Set $S=U^\top HU$ and $J_U=U^\top JU$. The inverse square root of the restricted metric and the normalized old frame satisfy
\begin{align*}
 (U^\top Q(h)U)^{-1/2}
 &=I_p-\tfrac12hS+h^2(\tfrac38S^2-\tfrac14J_U)+O(h^3),\\
 \widetilde U
 &=U\bigl\{I_p-\tfrac12hS+h^2(\tfrac38S^2-\tfrac14J_U)\bigr\}+O(h^3).
\end{align*}
Substituting into the overlap gives
\[
 U(h)^\top Q(h)\widetilde U
 =I_p+hX+h^2Y_{\rm RP}+O(h^3),
\]
where
\begin{align}
 X&=V^\top U+\tfrac12S=-\omega(0),\label{eq:rpx}\\
 Y_{\rm RP}&=\tfrac12W^\top U+V^\top HU+\tfrac14J_U
              -\tfrac12V^\top US-\tfrac18S^2.\label{eq:rpsecond}
\end{align}
By \cref{eq:constraint1}, $X^\top=-X$, so polar normalization leaves the first order coefficient unchanged. The second order coefficient determines the discrepancy from exact transport.

\subsection{Geodesic overlap and polar projection}

The left endpoint whitening reduces \cref{eq:geodesicmap} to $T_{\rm geo}=Q(h)^{-1/2}$. Its matrix binomial expansion is
\begin{equation}\label{eq:geoexpansion}
 T_{\rm geo}
 =I_n-\tfrac12hH+h^2(\tfrac38H^2-\tfrac14J)+O(h^3).
\end{equation}
For comparison, the exact ambient propagator solves
\[
 \dot T=-\tfrac12Q^{-1}\dot Q\,T,
 \qquad T(0)=I_n,
\]
and differentiation at the left endpoint gives
\[
 \dot T(0)=-\tfrac12H,
 \qquad \ddot T(0)=-\tfrac12J+\tfrac34H^2.
\]
Its Taylor expansion therefore agrees with \cref{eq:geoexpansion} through order two. To pass from ambient transport to the represented subspace, expand the overlap as
\[
 M(h)=U(h)^\top Q(h)T_{\rm geo}U
     =I_p+hX+h^2Y+O(h^3),
\]
with
\begin{align}
 X&=V^\top U+\tfrac12U^\top HU=-\omega(0),\label{eq:xdef}\\
 Y&=\tfrac12W^\top U+\tfrac12V^\top HU
    -\tfrac18U^\top H^2U+\tfrac14U^\top JU.\label{eq:ydef}
\end{align}
Since $X^\top=-X$, the positive factor used in the polar normalization satisfies
\begin{align*}
 M(h)^\top M(h)
 &=I_p+h^2(Y+Y^\top-X^2)+O(h^3),\\
 (M(h)^\top M(h))^{-1/2}
 &=I_p-\tfrac12h^2(Y+Y^\top-X^2)+O(h^3).
\end{align*}
Multiplying by $M(h)$ yields the expansion
\begin{equation}\label{eq:polarexpansion}
 \polar(M)
 =I_p+hX+h^2\{\Skew(Y)+\tfrac12X^2\}+O(h^3).
\end{equation}
Differentiating the connection form in \cref{eq:omega} gives
\begin{equation}\label{eq:omegadotexpanded}
 \begin{aligned}
 \dot\omega(0)
 ={}&V^\top V+U^\top W+\tfrac32U^\top HV\\
    &+\tfrac12V^\top HU+\tfrac12U^\top JU.
 \end{aligned}
\end{equation}
Its derivative is skew symmetric because $\omega$ is skew symmetric. Taking the skew part, equivalently using \cref{eq:constraint2}, gives
\begin{equation}\label{eq:skewy}
 \begin{aligned}
 \dot\omega(0)
 &=\tfrac12(U^\top W-W^\top U)
   +\tfrac12(U^\top HV-V^\top HU),\\
 \Skew(Y)&=-\tfrac12\dot\omega(0).
 \end{aligned}
\end{equation}
The exact coefficient propagator has the Taylor expansion
\[
 \Psi(h,0)
 =I_p-h\omega(0)
  +\tfrac12h^2\{\omega(0)^2-\dot\omega(0)\}+O(h^3).
\]
Substituting \cref{eq:xdef,eq:skewy} into \cref{eq:polarexpansion} reproduces this expansion, proving the local error $R_{\rm geo}(h,0)-\Psi(h,0)=O(h^3)$.

\subsection{The Procrustes defect coefficient}

Subtracting \cref{eq:ydef} from \cref{eq:rpsecond} isolates the term that differs between the two rules:
\[
 Y_{\rm RP}-Y
 =\tfrac12V^\top(I_n-UU^\top)HU
  +\tfrac18\{U^\top H^2U-S^2\}.
\]
The expression in braces is symmetric and therefore vanishes under $\Skew$. In the current whitened coordinates, define
\[
 B=(I_n-UU^\top)HU,
 \qquad N=(I_n-UU^\top)(V+HU/2).
\]
The skew part becomes
\begin{equation}\label{eq:defectproof}
 \begin{aligned}
 \Skew(Y_{\rm RP}-Y)
 &=\tfrac14\bigl\{V^\top(I_n-UU^\top)HU
                 -U^\top H(I_n-UU^\top)V\bigr\}\\
 &=\tfrac14(N^\top B-B^\top N).
 \end{aligned}
\end{equation}
The common term $X^2/2$ in \cref{eq:polarexpansion} cancels in the comparison. Undoing the constant whitening replaces the Euclidean inner products by their $Q$ counterparts, giving
\[
 \mathcal D(t)=\tfrac14\bigl(N(t)^\top Q(t)B(t)-B(t)^\top Q(t)N(t)\bigr).
\]
Together with the geodesic expansion, this proves all three local statements in \cref{eq:rpexactdefect,eq:geolocalorder,eq:methoddifference}.

\subsection{Uniform remainders and admissible steps}

The spectral and derivative bounds in \cref{thm:orders} place every whitened step in a common compact neighborhood on which the required matrix functions have bounded derivatives. Each overlap is $I_p+O(h)$ uniformly in the left endpoint, so for some uniform $C$ and all sufficiently small steps,
\[
 \sigma_{\min}(M(h))\geq1-\|M(h)-I_p\|_2
                       \geq1-Ch>\tfrac12.
\]
The same argument applies to the risk weighted overlap, proving that both polar factors are defined on a uniform range of step sizes. Taylor's theorem then gives uniform local remainders. Orthogonal changes of endpoint frames preserve overlap singular values and transport error norms, so independently computed frames need not themselves have a smooth orientation.

\subsection{Accumulation on nonuniform partitions}

Let $h_k=t_{k+1}-t_k$, let $R_k$ be either numerical update, and let $\Psi_k$ be the exact coefficient propagator. If $e_k$ denotes numerical minus exact coefficients, then
\[
 e_{k+1}=R_ke_k+(R_k-\Psi_k)c(t_k).
\]
Orthogonality of $R_k$ and preservation of $\|c(t_k)\|=\|c_0\|$ imply
\[
 \|e_{k+1}\|\leq\|e_k\|+Ch_k^{q+1}\|c_0\|.
\]
Here $q=1$ for risk weighted Procrustes in general, and $q=2$ for the geodesic rule or for Procrustes when $\mathcal D$ vanishes along the path. Iteration and the partition identity give
\[
 \sum_k h_k^{q+1}
 \leq h_{\max}^q\sum_kh_k
 =Th_{\max}^q,
\]
and hence
\begin{equation}\label{eq:globalaccumulation}
 \max_k\|e_k\|\leq\|e_0\|+CT\|c_0\|h_{\max}^q.
\end{equation}
Since $U(t_k)^\top Q(t_k)U(t_k)=I_p$, the corresponding asset error satisfies $\|U(t_k)e_k\|_{Q(t_k)}=\|e_k\|$. Thus the same bound holds in the risk norm, completing the usual accumulation argument for orthogonal propagators \citep{harland2023,zhang2026} after the explicit metric dependent local comparison above.
\hfill$\square$

\section{Stability constants and projector control}\label{app:stability}

\subsection{From spectral projectors to aligned frames}

The frame term in \cref{eq:stability} can be linked to primitive estimation errors. Let the represented subspace be the spectral cluster associated with an interval of a symmetric whitened operator $L_t$, separated from the remaining spectrum by $\delta$. If
\begin{equation}\label{eq:gapcondition}
 \sup_{t\leq T}\norm[2]{\widehat L_t-L_t}<\delta/2,
\end{equation}
the Davis-Kahan theorem gives
\begin{equation}\label{eq:projectorbound}
 \sup_{t\leq T}\norm[2]{\widehat P_t^{\rm E}-P_t^{\rm E}}
 \leq \frac{2}{\delta}
 \sup_{t\leq T}\norm[2]{\widehat L_t-L_t},
\end{equation}
where the superscript denotes Euclidean projectors in whitened coordinates. Orthogonal Procrustes alignment then yields
\begin{equation}\label{eq:framebound}
 \norm[2]{\widehat U_tR_t-U_t}
 \leq \frac{\sqrt2}{\sqrt\alpha}\norm[2]{\widehat P_t^{\rm E}-P_t^{\rm E}}
 +\norm[2]{\widehat Q_t^{-1/2}-Q_t^{-1/2}}.
\end{equation}
For spectra in $[\alpha,\beta]$, functional calculus gives the operator Lipschitz bound
\begin{equation}\label{eq:sqrtbound}
 \norm[2]{\widehat Q^{-1/2}-Q^{-1/2}}
 \leq \frac{1}{2\alpha^{3/2}}\norm[2]{\widehat Q-Q}.
\end{equation}
Combining \cref{eq:projectorbound,eq:framebound,eq:sqrtbound} gives \cref{eq:daviskahan}.

\subsection{Connection error and temporal derivatives}

The connection error depends on both the geometry and its temporal derivatives, as seen by expanding \cref{eq:omega}:
\begin{multline}\label{eq:connectionerror}
 \norm[2]{\widehat\omega-\omega}
 \leq C_1\norm[2]{\widehat U-U}
 +C_2\norm[2]{\partial_t\widehat U-\partial_tU}\\
 +C_3\norm[2]{\widehat Q-Q}
 +C_4\norm[2]{\partial_t\widehat Q-\partial_tQ},
\end{multline}
where the constants depend only on $\alpha$, $\beta$, and uniform bounds for the true and estimated frames and derivatives. Thus covariance consistency without temporal derivative control is insufficient for connection consistency. This is why the empirical method regularizes updates and why gap collapse dates fall outside the smooth theory.

\subsection{Accumulation of discrete estimation errors}

For discrete endpoint transport, let $\mathcal R_k$ and $\widehat{\mathcal R}_k$ be the exact and fitted orthogonal coefficient maps. If $\norm{\widehat{\mathcal R}_k-\mathcal R_k}\leq\eta_k$, then $e_k=\widehat c_k-c_k$ satisfies
\begin{align*}
 e_{k+1}&=\widehat{\mathcal R}_ke_k
          +(\widehat{\mathcal R}_k-\mathcal R_k)c_k,\\
 \|e_{k+1}\|&\leq\|e_k\|+\eta_k\|c_0\|.
\end{align*}
Iterating this inequality gives
\begin{equation}\label{eq:discretestability}
 \norm{\widehat c_K-c_K}
 \leq \norm{\widehat c_0-c_0}+\norm{c_0}\sum_{k=0}^{K-1}\eta_k.
\end{equation}
Unlike a generic product bound, no exponential factor appears because both propagators are norm preserving.

\section{Experimental specification}\label{app:protocol}

The first manufactured problem uses four asset coordinates and a subspace of dimension two, with an analytic coefficient path. A fourth order Runge--Kutta calculation verifies numerical integration independently of the endpoint methods, and smooth orthogonal changes of frame test gauge covariance and the motion decomposition. The second problem uses six asset coordinates and a subspace of dimension three. Its metric has rotating eigenspaces and unequal varying eigenvalues, with a diagonal path providing the commuting control. DOP853 solves the continuous coefficient equation at relative tolerance $2\times10^{-12}$. Refinements compare terminal errors in the risk metric. The rank two construction in \cref{prop:sharp} separately uses the exact identity propagator, so its reference has neither integration error nor a preference for a numerical method.

Both financial panels follow the uninterrupted availability rule of Section~4. At decision index $t$, the covariance $Q_t$ is the Ledoit--Wolf estimate from the preceding 504 observations plus $10^{-10}I_n$. If $E_t$ contains the leading five covariance eigenvectors for industries or eight for size and value, define
\begin{equation}\label{eq:empiricalframe}
 M_t=Q_t^{-1}-\frac{Q_t^{-1}\mathbf1\mathbf1^\top Q_t^{-1}}
 {\mathbf1^\top Q_t^{-1}\mathbf1},
 \qquad
 U_t=M_tE_t(E_t^\top M_t^\top Q_tM_tE_t)^{-1/2}.
\end{equation}
The columns of $U_t$ are risk orthonormal and have zero total weight. The signal compounds 252 daily returns ending 21 observations before the decision. Its components are clipped at the cross sectional median plus or minus four median absolute deviations, and their cross sectional mean is removed. The local active direction is $U_tU_t^\top s_t$, rescaled to annual volatility $5\%$ using 252 observations per year. An equal weight anchor makes the full portfolio fully invested, and radial contraction of the active component imposes a gross exposure cap of two. The cap can reduce active risk below the target. For completeness, ambient projection and the Euclidean coefficient alignment are
\begin{equation}\label{eq:additionalbenchmarks}
 v_{\rm amb}=U_1U_1^\top Q_1v_0,
 \qquad R_{\rm E}=\polar(U_1^\top U_0),
\end{equation}
while the metric based rules are given in \cref{eq:rp,eq:meanoverlap,eq:symmetricprocrustes}. The carried state is the preceding implemented active target, expressed in its stored frame and metric, rather than the buy and hold portfolio immediately before the next trade. Each update transports that stored target, blends it with the new local direction, rescales active risk, and applies the gross cap. The primary weight on the new target is $0.25$, and the comparison grid is $0.10$, $0.25$, $0.50$, and $0.75$. Local refitting instead discards the stored target, while all methods start from the same equal weight allocation and share the signal, constraints, and initialization. Rebalancing occurs every 21 observations, with 5 and 63 observations used only for sensitivity analysis. Endpoint overlaps with a smallest singular value below $10^{-10}$ times the largest are rejected rather than assigned a polar completion, and no such rejection occurred in the reported runs.

Between decisions, holdings are self financing and weights evolve with realized returns. Turnover at a decision is one half of the absolute weight difference between the desired target and the actual drifted portfolio. The specified cost convention deducts the fraction $c\,\mathrm{TO}$ from wealth before applying the first daily return of the new holding interval, with $c=0.001$ for industries and $c=0.0015$ for size and value. Thus the first net gross return factor is $(1-c\,\mathrm{TO})(1+r)$, and all later daily factors in the interval are $1+r$. Certainty equivalent is computed as
\begin{equation}\label{eq:ceempirical}
 \mathrm{CE}=252\,\bar r-\frac{3}{2}\,252\,s_r^2,
\end{equation}
where $\bar r$ and $s_r^2$ are the arithmetic mean and sample variance of daily net returns. Annualized turnover is total realized turnover times 252 divided by the number of evaluated daily returns, including the final incomplete holding interval. The paired bootstrap resamples aligned daily strategy returns in blocks of three rebalancing intervals and aligned decision level turnover and tracking differences in blocks of six decisions, using 500 replications. The composition diagnostic uses 1,000 replications with disjoint starting intervals and a reference formed from geometric mean updates at five observation intervals over the same 21 or 63 observation horizon. The reported intervals are conditional on the selected panels and protocol, excluding uncertainty from data revisions and retrospective sample selection, and are unadjusted for comparisons across the sensitivity grid. Deterministic implementation tests verify weight drift, the cost convention, calendar continuity, and the treatment of repeated eigenvalues.

\end{document}